\documentclass[jcp,aps,twocolumn,showpacs,amsmath,amssymb,superscriptaddress]{revtex4-1}
\usepackage{graphicx}
\usepackage{braket}
\usepackage{amsmath}
\usepackage{bm}

\begin{document}
\title{Non-Markovian finite-temperature 
two-time correlation functions of system operators: beyond the quantum regression theorem}
\author{ Hsi-Sheng Goan}
\email[Corresponding author: ]{goan@phys.ntu.edu.tw}
\author{Po-Wen Chen}
\author{Chung-Chin Jian}
\affiliation{Department of Physics  and Center for Theoretical Sciences, National Taiwan University, Taipei 10617, Taiwan}
\affiliation{Center for Quantum Science and Engineering, National Taiwan University, Taipei 10617, Taiwan}
\begin{abstract}

An extremely useful evolution equation that allows
systematically calculating the two-time correlation functions (CF's)
of system operators for non-Markovian open (dissipative) quantum systems
is derived. The derivation is based on perturbative quantum master
equation approach, so non-Markovian open quantum system models that are
not exactly solvable can use our derived evolution equation to easily
obtain their two-time CF's of system operators, valid to second order in
the system-environment interaction. 
Since the form and nature of the Hamiltonian are not specified in our
derived evolution equation,  
our evolution equation is applicable for
bosonic and/or fermionic environments and can be applied to a wide range
of system-environment models with 
any factorized (separable) system-environment initial states (pure or mixed). 
When applied to a general model of a system coupled to a
finite-temperature bosonic environment with a  
system coupling operator $L$ in the system-environment interaction
Hamiltonian, the resultant evolution equation is
valid for both $L=L^\dagger$ and $L\neq L^\dagger$ cases, in contrast
to those evolution equations valid only for $L=L^\dagger$ case in the literature.
The derived equation that 
generalizes the quantum regression theorem (QRT) 
to the non-Markovian case will have broad applications in
many different branches of physics. 
We then give conditions on which the QRT holds in the weak
system-environment coupling case, and apply the derived evolution
equation to a problem of a two-level system (atom) coupled to a
finite-temperature bosonic environment (electromagnetic fields) with
$L\neq L^\dagger$.   

\end{abstract}

\pacs{03.65.Ca, 03.65.Yz, 05.30.-d, 42.50.Lc}

\maketitle

\section{Introduction}
In many fields of modern sciences, one has to deal with open quantum
systems in contact with their quantum surroundings or environments
(reservoirs or baths) \cite{Scully97, Carmichael99,Barnett02,Gardiner00,Milburn08,Paz01,Breuer02,Grabert88,Weiss08}. 
Most often, one concerns with only the system dynamics and the key
quantity is the reduced system density matrix $\rho(t)$ defined as the
partial trace of the total system-plus-reservoir density operator
$\rho_T(t)$ over the reservoir degrees of freedom; i.e., $\rho(t)={\rm
  Tr}_R[\rho_T(t)]$. The evolution equation of the reduced density
matrix is governed by the reduced Liouville equation or called the
quantum master equation that can be Markovian or non-Markovian.

Two-time (multi-time) correlation functions (CF's) 
of an open quantum system 
are important physical quantities. They can provide significant
information about the system, whereas the single-time expectation values
can not.
For example, the two-time CF's of the
electromagnetic field emitted by an atom are required for calculating
the fluorescence spectrum \cite{Scully97,Carmichael99,Barnett02,Gardiner00,Milburn08}. 
The two-time CF's of the 
number of emitted photons give the information about the photon
statistics and describe the behavior of photon bunching and
anti-bunching\cite{Scully97,Carmichael99,Barnett02,Gardiner00,Milburn08}. The
two-time CF's of the electric current through
nanostructure devices are useful in the study of the transport
properties of current fluctuations and noise spectrum \cite{Blanter00,Clerk10}. 
In the Markovian case, an extremely useful procedure to calculate the
two-time (multi-time) CF's for open (dissipative) quantum systems
is the so-called quantum regression theorem (QRT) \cite{Scully97,Carmichael99,Barnett02,Gardiner00,Milburn08} that gives
 a direct relation between the time evolution equation of the
 single-time expectation values and that of their corresponding
 two-time (multi-time) CF's. 
So knowing the time evolution of the reduced density matrix of
the system allows one to calculate all of the single-time expectation
values and two-time (multi-time) CF's in the Markovian case.
This is, however, no longer true in the non-Markovian case.
For non-Markovian open (dissipative) quantum systems,
the QRT is not valid in 
general \cite{Grabert88,Weiss08,Grabert82,Talkner95,Ford96,Ford99,Lax00,Ford00}.
Although it is commendable to use the exact procedures \cite{Grabert88,Weiss08,Ford96,Ford99}
to calculate
directly the non-Markovian two-time CF's,
not too many problems can be exactly worked out in this way.
It is thus important that a procedure similar to the QRT can
be developed and be applied to calculate the two-time CF's
perturbatively 
for the non-Markovian open (dissipative) quantum systems.

Recently, by using the stochastic Schr\"{o}dinger equation approach
and the Heisenberg equation of system operator method,  
Alonso and de Vega derived \cite{Alonso05,Vega06,Alonso07} the evolution equations 
of the two-time (multi-time) CF's of the system operators 
for a general model of a quantum system coupled to a
bosonic environment with a system coupling operator $L$ in the
system-environment interaction Hamiltonian. 
The evolution equations,
valid to second order in system-environment coupling strength,  
were applied to calculate the emission spectra of a two-level atom
placed in a structured non-Markovian environment, i.e, electromagnetic
fields in a photonic band-gap material \cite{Vega08}.  
In the photonic band-gap material, the correlation function of the
electromagnetic field (environment) is highly non-Markovian,
particularly within the edges of the bands, so the QRT is not valid
even when the atom-environment interaction strength is weak.  
Although the evolution equations, derived in
Refs.~\cite{Alonso05,Vega06,Alonso07} (Eq.~(6) in
Ref.~\cite{Alonso05}, Eq.~(31) in Ref.~\cite{Vega06} and Eq.~(60) in
Ref.~\cite{Alonso07}), are very useful to calculate the time evolution of
the {\em two-time (multi-time)} CF's of the system observables, 
the derivations of the evolution equations 
were presented only for an environment at zero temperature and only for a
system state in an initial pure state.  
In Ref.~\cite{Vega06}, it was mentioned that it is possible to use the
reduced stochastic system propagator that corresponds to an initial
state of the environment different from the vacuum to evaluate the
single-time expectation values and multi-time CF's with more general
initial conditions.  But they derived only a master equation that is
conditioned on initial bath states and is capable of evaluating just the
{\em single-time} expectation values of system observables for general
initial conditions, both for an initial pure state and mixed state
\cite{Vega06}.  
So strictly speaking, the two-time evolution equations derived 
in Refs.~\cite{Alonso05,Vega06,Alonso07} is applicable only for a
zero-temperature environment. 
However, these equations
were used to calculate the
two-time CF's of system observables of dissipative spin-boson models
at finite temperatures \cite{Alonso05,Vega06,Alonso07}.    
This is possible only for the dissipative spin-boson models 
with Hermitian system coupling
operators, $L=L^\dagger$
(see also the discussions in subsection \ref{sec:comparison}).

In this paper, we use another commonly used technique, the quantum master equation approach \cite{Scully97,Carmichael99,Barnett02,Gardiner00,Milburn08,Paz01,Breuer02},  
to derive in the weak system-environment coupling limit an evolution
equation of the two-time CF's of the system operators 
for non-Markovian open quantum systems
in finite-temperature environments
{\em for any factorized (separable) system-environment initial states (pure or mixed)}. 
This quantum master equation approach, different from those in
Refs.~\cite{Alonso05,Vega06,Alonso07}, allows us to explicitly point out an
important nonlocal environment (bath) memory term that vanishes in the
Markovian case but makes the evolution equation deviate from the QRT
in general cases.   
Since the form and nature of the Hamiltonian are not specified in our
derived evolution equation,  
the evolution equation is applicable for  
bosonic and/or fermionic environments and can be applied to a wider range
of system-environment models. 
When applied to a general model of a system coupled to a
finite-temperature bosonic environment, 
our resultant two-time evolution equation is
{\em valid for both Hermitian  
and non-Hermitian system coupling operator cases}.
This is in contrast to the two-time evolution equations derived 
in Refs.~\cite{Alonso05,Vega06,Alonso07} valid
only for the finite-temperature
bosonic environment case
with a Hermitian system coupling operator $L=L^\dagger$. 
Our derived equation that generalizes the QRT to the non-Markovian
case will have broad applications in
many different branches of physics. 

The paper is organized as follows. 
In Sec.~\ref{sec:derivation}, we derive the non-Markovian
finite-temperature evolution equations 
for one-time expectation values and 
two-time CF's for a general
system-environment thermal model. 
We then apply the derived equations (valid to second order in
the system-environment interaction) 
to a thermal bosonic environment and work out the explicit forms of
the evolution equations in terms of system coupling operators in
Sec.~\ref{sec:bosonic}. 
The evolution equations
for a non-Markovian bosonic environment has been presented in
Ref.~\cite{Goan10} without any derivation. In this paper, the detailed
derivation is given.
We also 
discuss the connection of our derived two-time 
evolution equations with those presented in 
Refs.~\cite{Alonso05,Vega06,Alonso07}
and give conditions on which the QRT may hold in the 
weak system-environment coupling case. In Sec.~\ref{sec:spin-boson}, we 
apply the newly derived equation to a problem of a two-level system (atom) coupled to a bosonic environment (electromagnetic fields),
in which the system coupling operator is non-Hermitian, i.e., $L\neq L^\dagger$.  A short conclusion is given in Sec.~\ref{sec:conclusion}.

\section{Derivations of evolution equations}\label{sec:derivation}

\subsection{Time-nonlocal and  time-local  quantum master equations}

Let us consider a general total Hamiltonian of the system 
and reservoir (environment) as
\begin{equation}
H =H_{S}+H_{I}+H_{R}, 
\label{Hamiltonian}
\end{equation}
where $H_{S}$ and $H_{R}$ are the Hamiltonian for system and reservoir, respectively, and $H_{I}$ is the interaction between the system and reservoir.
It is convenient to go to the interaction picture in which 
\begin{equation}
{d\tilde{\rho}_{T}(t)}/{dt} 
=-({i}/{\hbar})[ \tilde{H}_{I}\left( t\right) ,\tilde{\rho}_{T}(t)
],  \label{rhoT_int_eom}
\end{equation}%
where
$\tilde{H}_{I}\left( t\right)= e^{iH_{0}t/\hbar
}H_{I}e^{-iH_{0}t/\hbar }$
is explicitly time-dependent,
$\tilde{\rho}_{T}(t)= e^{iH_{0}t/\hbar }{\rho}_{T}(t)e^{-iH_{0}t/\hbar }$
is the total 
density matrix operator 
at time $t$ in the interaction picture and 
$H_{0}=H_{S}+H_{R}$.
Then the Liouville equation of the total density matrix $\rho _{T}(t)$ in the Schr\"{o}dinger picture becomes
\begin{equation}
   \frac{d\rho _{T}(t)}{dt}=-\frac{i}{\hbar}\left[ H_0,\rho _{T}(t)\right]
+e^{-iH_{0}t/\hbar}\, \frac{d\tilde{\rho}_{T}(t)}{dt}\, e^{iH_{0}t/\hbar }.
\label{rh0T_relation}
\end{equation}

One can integrate Eq.~(\ref{rhoT_int_eom}) formally to obtain 
\begin{equation}
\tilde{\rho}_{T}\left( t\right) =\tilde{\rho}_{T}\left( 0\right) 
-\frac{i}{\hbar }\int\nolimits_{0}^{t}dt^{\prime }\left[ \tilde{H}_{I}\left(
t^{\prime }\right) ,\tilde{\rho}_{T}\left( t^{\prime }\right) \right] ,
\label{rhoT_int_iteration}
\end{equation}%
and substitute for $\tilde{\rho}_{T}\left( t\right) $ inside the commutator
in Eq.~(\ref{rhoT_int_eom}). 
The resultant equation is 
\begin{eqnarray}
\frac{d\tilde{\rho}_{T}(t)}{dt} &=&-\frac{i}{\hbar }\left[ \tilde{H}_{I}\left(
t\right) ,\tilde{\rho}_{T}\left( 0\right) \right]  \nonumber \\
&&-\frac{1}{\hbar ^{2}}\int\nolimits_{0}^{t}dt^{\prime }\left[ \tilde{H}%
_{I}\left( t\right) ,\left[ \tilde{H}_{I}\left( t^{\prime }\right) ,\tilde{%
\rho}_{T}\left( t^{\prime }\right) \right] \right] .  
\label{tilderho}
\end{eqnarray}%
Equation (\ref{tilderho}) is still exact. 
Suppose initially   $\rho_{T}(0)=\tilde{\rho}_{T}(0)=\rho(0)\otimes R_0$, 
where $R_0=\exp(-H_R/k_BT)/{\rm Tr}_R[\exp(-H_R/k_BT)]$ is the initial 
thermal reservoir density operator. 
Then after tracing over the reservoir, Eq.~(\ref{tilderho}) gives 
the master equation of the reduced density matrix in the interaction picture 
\begin{eqnarray}
\frac{d\tilde{\rho}(t)}{dt} 
=-\frac{1}{\hbar ^{2}}{\rm Tr}_R\int\nolimits_{0}^{t}dt^{\prime }
\left[ \tilde{H}_{I}\left( t\right) ,\left[ \tilde{H}_{I}
\left( t^{\prime }\right) ,\tilde{\rho}_{T}\left( t^{\prime }\right) \right] 
\right],
\label{tilderho_reduced}  
\end{eqnarray} 
where we have 
considered a class of system-reservoir interaction Hamiltonian models 
for the thermal reservoir such that 
\begin{equation}
{\rm Tr}_R[\tilde{H}_{I}(t) R_0]=0
\label{traceless_1st_order}
\end{equation}
to eliminate the first term in Eq.~(\ref{tilderho}).

The perturbative non-Markovian open quantum system theory may be
categorized into two classes:  the time-nonlocal (or time-convolution)
\cite{Breuer02,Meier99,Breuer99,Breuer01,Yan98,Schroder06,Ferraro09} and the
time-local (or time-convolutionless)
\cite{Breuer02,Paz01,Breuer99,Breuer01, Yan98,Schroder06,Ferraro09,Shibata77,Kleinekathofer04,Liu07,Sinayskiy09,Mogilevtsev09,Haikka10,Ali10}
methods. This has been discussed extensively in
the literature \cite{Breuer02,Breuer99,Schroder06,Ferraro09}.  
The environment or reservoir by definition is large and contains
many degrees of freedom so that the influence of the system on the
reservoir is small in the weak system-environment coupling case. 
As a consequense, to second order in system-environment interaction,
the total density operator 
on the right hand side of Eq.~(\ref{tilderho_reduced}) 
can be approximated to 
an uncorrelated (factorized) state as 
$\tilde{\rho}_T(t')=\tilde{\rho}(t')\otimes R_0+ {\cal O}(\tilde{H}_I)$ 
\cite{Carmichael99} since the products of 
two interaction Hamiltonians $\tilde{H}_I$'s 
appear already there. 
So in many textbooks \cite{Scully97,Carmichael99,Barnett02}, 
the replacement of $\tilde{\rho}_T(t')$ with $\tilde{\rho}(t')\otimes R_0$ 
in Eq.~(\ref{tilderho_reduced}) is performed 
under the so-called Born approximation. 
One then obtains \cite{Scully97,Carmichael99,Barnett02,Breuer02,Meier99,Breuer99,Breuer01, Yan98,Schroder06,Ferraro09}
\begin{eqnarray}
\frac{d\tilde{\rho}(t)}{dt} 
=-\frac{1}{\hbar ^{2}}{\rm Tr}_R\int\nolimits_{0}^{t}dt^{\prime }
\left[ \tilde{H}_{I}\left( t\right) ,\left[ \tilde{H}_{I}
\left( t^{\prime }\right) ,\tilde{\rho}\left( t'\right) \otimes R_0 \right] 
\right]. \nonumber \\
\label{tilderho_reduced_pert}  
\end{eqnarray} 
Note that Eq.~(\ref{tilderho_reduced_pert}) is in a
form of delayed integro-differential equation and is thus a time-nonlocal
master equation.
However, it can also be shown that another systematically 
perturbative non-Markovian master
equation that is local in time
\cite{Breuer02,Paz01,Breuer99,Breuer01, Yan98,Schroder06,Ferraro09,Shibata77,Kleinekathofer04,Liu07,Sinayskiy09,Mogilevtsev09,Haikka10,Ali10}
can be derived from the time-convolutionless
projection operator formalism \cite{Breuer02,Breuer99,Breuer01, Yan98,Schroder06,Shibata77} or from the iteration
expansion method \cite{Paz01}.  
Under the similar assumption of the factorized initial system-reservoir
density matrix state and the use of Eq.~(\ref{traceless_1st_order}), the
second-order time-convolutionless master equation in the interaction
picture can be obtained as 
\cite{Breuer02,Paz01,Breuer99,Breuer01, Yan98,Schroder06,Ferraro09,Shibata77,Kleinekathofer04,Liu07,Sinayskiy09,Mogilevtsev09,Haikka10,Ali10}
\begin{eqnarray}
\frac{d\tilde{\rho}(t)}{dt} 
=-\frac{1}{\hbar ^{2}}{\rm Tr}_R\int\nolimits_{0}^{t}dt^{\prime }
\left[ \tilde{H}_{I}\left( t\right) ,\left[ \tilde{H}_{I}
\left( t^{\prime }\right) ,\tilde{\rho}\left( t\right) \otimes R_0 \right] 
\right]. \nonumber \\
\label{time_convolutionless_ME}  
\end{eqnarray} 
We note here that obtaining the time-convolutionless 
non-Markovian master equation perturbatively
up to only second order 
in the interaction Hamiltonian 
using the time-convolutionless projection operator technique 
\cite{Breuer02,Breuer99,Breuer01}
is equivalent to obtaining it 
by replacing $\tilde{\rho}(t')$ with $\tilde{\rho}(t)$ 
in Eq.~(\ref{tilderho_reduced_pert}) \cite{Breuer02,Paz01,Breuer99,Breuer01, Yan98,Schroder06,Ferraro09,Shibata77,Kleinekathofer04,Liu07,Sinayskiy09,Mogilevtsev09,Haikka10,Ali10}.
One may be tempted to think that the second-order 
time-nonlocal master equation (\ref{tilderho_reduced_pert}) is
more accurate than the second-order time-local (time-convolutionless)
master equation (\ref{time_convolutionless_ME}) since besides the Born approximation, the  (first) Markovian
approximation of replacing $\tilde{\rho}(t')$ with $\tilde{\rho}(t)$
in Eq.~(\ref{tilderho_reduced_pert}) 
seems to be an additional approximation made on the
time-local master equation. But it was shown that this may not be the
case. In many examples \cite{Breuer02,Breuer99,Breuer01, Yan98,Ferraro09,Kleinekathofer04}, the time-convolutionless approach
works better than the time-nonlocal approach when the exact dynamics
is used to test the perturbative non-Markovian theory based on these
two approaches.
Here we will consider the second-order non-Markovian
time-convolutionless (time-local) evolution equation in our derivation.

\subsection{Quantum regression procedure}
\label{sec:QRT}
Quantum regression procedure or 
QRT indicates that for (Markovian) open quantum systems, the equations of motion
(or evolution equations) for single-time expectation values of system operators are also the
equations of motion for two-time (multi-time) CF's.
Formally, the single-time or one-time expectation values of system operators 
can be written as    
\begin{equation}
\left\langle A\left( t_{1}\right) \right\rangle 
={\rm Tr}_{S\otimes R}\left[A\left( t_{1}\right) \rho_{T}\left( 0\right)\right]
={\rm Tr}_{S\otimes R}\left[  A\left( 0\right)\rho_{T}\left( t_1\right) \right],  
\label{A1}
\end{equation}
where $A(t_1)$ represents a general system Heisenberg operator(s), 
$\rho_T(t_1)$ is the Schr\"{o}dinger total density matrix operator
at time $t_1$, and the subscript of ${S\otimes R}$ attached to the
trace symbol of ${\rm Tr}$ indicates a trace over the Hilbert space of
the total composite system \cite{direct_product}. 
The evolution equation of the one-time expectation values 
of system operators can then be obtained as  
\begin{eqnarray}
{d\left\langle A\left( t_{1}\right) \right\rangle }/{dt_{1}}
&=&{\rm Tr}_{S\otimes R}\left[ A ({d\rho_T(t_1)}/{dt_1})\right] 
\label{1time_evol}\\
&=&{\rm Tr}_{S}\left[ A({d\rho(t_1)}/{dt_1})\right],
\label{1time}
\end{eqnarray}
where we have used the fact that $A=A(0)$ is a pure system operator and  
${\rm Tr}_R[{d\rho_T(t_1)}/{dt}]={d\rho(t_1)}/{dt}$.
For the two-time CF with $t_1>t_2$, one has 
\begin{eqnarray}
\left\langle A\left( t_{1}\right) B\left( t_{2}\right) \right\rangle
&=&{\rm Tr}_{S\otimes R}\left[A\left( t_{1}\right) B\left(
t_{2}\right)  \rho _{T}\left( 0\right) \right]
\nonumber \\
&=&{\rm Tr}_{S\otimes R}\left[ A e^{-iHt/\hbar}B \rho _{T}\left( t_{2}\right)
e^{iHt/\hbar}\right], 
\label{AB12}
\end{eqnarray}
where $t=t_1-t_2$, and $A=A(0)$ and $B=B(0)$ are system operators. 
Let $\chi_T(0)=B \rho _{T}\left( t_{2}\right)$. 
Then the two-time CF (\ref{AB12}) becomes  $\left\langle A\left( t_{1}\right) B\left( t_{2}\right) \right\rangle
={\rm Tr}_{S\otimes R}\left[  A\chi_{T}\left( t\right) \right]$.  
It is then equivalent to
the expectation value of 
the operator $A$
with respect to the effective density matrix operator 
$\chi_T(t)=e^{-{iHt}/{\hbar}}B \rho _{T}\left( t_{2}\right)e^{{iHt}/{\hbar} }$
that satisfies the same Liouville evolution equation as $\rho_T(t)$
even though $\chi_T(t)$ may not be a proper density matrix 
(i.e., positive-definite trace-conservative operator).
The evolution equation of the two-time CF can be formally written as   
\begin{eqnarray}
{d\langle A\left( t_{1}\right) B(t_2)\rangle }/{dt_{1}}
&=&{\rm Tr}_{S\otimes R}\left[ A ({d\chi_T(t)}/{dt})\right] 
\label{2time_evol}\\
&=&{\rm Tr}_{S}\left[ A({d\chi(t)}/{dt})\right],
\label{2time}
\end{eqnarray}
where the relations of the reduced operator $\chi(t)={\rm Tr}_R[\chi_T(t)]$
and ${\rm Tr}_R[{d\chi_T(t)}/{dt}]={d\chi(t)}/{dt}$
have been used.
If the reduced master equations 
${d\chi(t)}/{dt}$ and  ${d\rho(t)}/{dt}$ had the same 
operator equation form, one might conclude that the structure and the form
of the evolution equation of the two-time CF would be the same as those of the single-time evolution equation and thus the QRT would apply.
In fact, it has been shown that 
the QRT is not valid in general \cite{Ford96,Ford99}, but 
the QRT or regression procedure is useful and correct for systems
where the coupling to reservoirs is weak 
and the Markovian approximation holds \cite{Carmichael99,Lax00,Ford00}.  
The main purpose of the present paper is to derive 
the non-Markovian finite-temperature evolution equation of the 
two-time system CF's using a quantum master equation approach, an
approach different from those in Refs.~\cite{Vega06,Alonso07}. 
Our equations, which are valid for both a Hermitian and a
non-Hermitian system coupling operators and thus generalize the
corresponding results in Refs.~\cite{Vega06,Alonso07}, can be used to
calculate the two-time CF's for any factorized (separable) 
system-reservoir initial
state and for any arbitrary temperature as long as the approximation
of the weak system-environment coupling still holds.

\subsection{Evolution equations in the weak system-environment  coupling limit}

Let us proceed to first derive perturbatively the explicit 
evolution equation of the
single-time expectation values in the  
non-Markovian case. 
Here we consider the second-order non-Markovian
time-convolutionless (time-local) evolution equation in our derivation. 
We wish to
obtain an evolution equation, $d\rho_T(t)/dt$, valid to second order
in system-environment interaction Hamiltonian, to substitute into
Eq.~(\ref{1time_evol}) for the single-time expectation values and into
Eq.~(\ref{2time_evol}) for the two-time CF's. 
It is convenient to first go to the interaction picture
and obtain a time-local (time-convolutionless) evolution equation of
the density matrix valid to that order. This can be achieved by the
substitution of $\tilde{\rho}_{T}(t')\to \tilde{\rho}_{T}(t)$ in the
second term on the right hand side of the equal sign of
Eq.~(\ref{tilderho})
\cite{Breuer02,Paz01,Breuer99,Breuer01, Yan98,Schroder06,Ferraro09,Shibata77,Kleinekathofer04,Liu07,Sinayskiy09,Mogilevtsev09,Haikka10,Ali10,Chen11}.
To go back to the Schr\"{o}dinger picture, we substitute the resultant
second-order equation obtained from Eq.~(\ref{tilderho}) into
Eq.~(\ref{rh0T_relation}) to obtain the evolution equation,
$d\rho_T(t)/dt$. By substituting this equation $d\rho_T(t)/dt$ valid
to second order in the interaction Hamiltonian into
Eq.~(\ref{1time_evol}), the evolution equation of the single-time
expectation values then consists of three terms. The second term involves
the first term on the right hand side of the equal sign of
Eq.~(\ref{tilderho}), and will vanish on the conditions that
$\rho_{T}(0)=\tilde{\rho}_{T}(0)=\rho(0)\otimes R_0$ and
${\rm Tr}_R[\tilde{H}_{I}(t) R_0]=0$ [Eq.~(\ref{traceless_1st_order})] 
are satisfied.   
As a result, we obtain up to second order in the interaction Hamiltonian
\begin{eqnarray}
\frac{d\left\langle A\left( t_{1}\right) \right\rangle }{dt_{1}}
&=&\frac{i}{\hbar }{\rm Tr}_{S\otimes R}\left(
[{H}_S , {A}] {\rho}_{T}(t_1)
\right)  \nonumber \\
&&+\frac{1}{\hbar^2 }\int_0^{t_1}d\tau {\rm Tr}_{S\otimes R} \nonumber \\
&&\quad
\left(\tilde{H}_I(\tau-t_1)[A,H_I]\rho_T(t_1) \right.
\nonumber \\
&&\quad\left.
+[H_I,A]\tilde{H}_I(\tau-t_1)\rho_T(t_1)\right) 
\label{1time_rhot}
\nonumber \\
&=&({i}/{\hbar}){\rm Tr}_{S\otimes R}\left( \{[{H}_S, {A}] \}(t_1) 
\rho_{T}(0)\right) \nonumber \\
&&+\frac{1}{\hbar^2 }\int_0^{t_1}d\tau {\rm Tr}_{S\otimes R} \nonumber\\
&&\quad
\left( \{\tilde{H}_I(\tau-t_1)[A,H_I]\}(t_1)\rho_T(0) \right. \nonumber\\
&&\quad\left.
+\{[H_I,A]\tilde{H}_I(\tau-t_1)\}(t_1)\rho_T(0)\right),
\label{1time_evol_eq}
\end{eqnarray}
where we have transformed from the Schr\"{o}dinger picture to the
Heisenberg picture in the second equal sign and 
$\{AB\}(t)\equiv \exp(iHt/\hbar) AB\exp(-iHt/\hbar)$.

Since $\chi_T(t)$ and $\rho_T(t)$ obey the same equations of 
Eqs.~(\ref{rh0T_relation}) and (\ref{tilderho}),
at first sight, one may think that the two-time evolution equations,
Eqs.~(\ref{2time_evol}) and (\ref{2time}), are similar to 
the single-time evolution equations, Eqs.~(\ref{1time_evol})  and
(\ref{1time}), and thus might be tempted to conclude that they have the
same form of the evolution equations.  
Indeed, by using Eqs.~(\ref{2time_evol}), (\ref{rh0T_relation}) 
and (\ref{tilderho}),  
the first and third terms of the resultant equation derived from
Eq.~(\ref{2time_evol}) are similar to the right-hand side of the single-time 
evolution equation (\ref{1time_evol_eq}) with the
replacement of 
$\rho_T(0)\to\chi_T(-t_2)=B(t_2)\rho_{T}(0)$ 
and with the change of the integration region from $[0,t_1]$ to
$[t_2,t_1]$. Then we obtain   
\begin{eqnarray}
&&\frac{i}{\hbar }{\rm Tr}_{S\otimes R}\left( \{[{H}_S, {A}] \}(t_1) 
B(t_2)\rho_{T}(0)\right) \nonumber \\
&&+\frac{1}{\hbar^2 }\int_{t_2}^{t_1}d\tau {\rm Tr}_{S\otimes R} \nonumber\\
&&\quad
\left( \{\tilde{H}_I(\tau-t_1)[A,H_I]\}(t_1)B(t_2)\rho_{T}(0) \right. \nonumber\\
&&\quad\left.
+\{[H_I,A]\tilde{H}_I(\tau-t_1)\}(t_1)B(t_2)\rho_{T}(0)\right).
\label{2time_evol_1_3}
\end{eqnarray}

However, a significant difference is that the expectation values for the second term does not vanish, i.e., 
\begin{equation}
(-i/\hbar){\rm Tr}_{S\otimes R}\left(Ae^{-iH_{0}t/\hbar}\left[\tilde{H}_{I}(t) ,\tilde{\chi}_{T}( 0)\right]e^{iH_{0}t/\hbar}\right)\neq 0,
\label{non-Markovian_1st_order}
\end{equation} 
in the non-Markovian case, where $t=t_1-t_2$ 
in Eq.~(\ref{non-Markovian_1st_order}). 
The reason can be understood as follows.
The interaction Hamiltonian $\tilde{H}_{I}(t_1-t_2)$ in
Eq.~(\ref{non-Markovian_1st_order}) involves the
environment operators in the time interval from $t_2$ to $t_1$, and 
the effective density matrix operator $\tilde{\chi}_T(0)$ can be written as 
$\tilde{\chi}_T(0)=\chi_T(0)=B\rho_T(t_2)=BU(t_2,0)\rho_T(0)U^{\dagger}(t_2,0)$,
where $U(t_2,0)=e^{-iHt_2/\hbar}$ is the Heisenberg evolution operator of the
total Hamiltonian from time $0$ to $t_2$. 
If the environment
is Markovian where the environment operator CF's at two
different times are $\delta$-correlated in time, then we may regard
that the environment operators in $\tilde{H}_{I}(t_1-t_2)$ are not
correlated with those in $U(t_2,0)$. So the trace over the environment degrees
of freedom for operator $\tilde{H}_{I}(t_1-t_2)$ and operator $U(t_2,0)$ can be
performed independently or separately. The trace of 
$\rho_T(t_2)=U(t_2,0)\rho_T(0)U^{\dagger}(t_2,0)$  over the environment
degrees of freedom
yields the reduced density matrix 
$\rho(t_2)={\rm Tr}_{R}[\rho_T(t_2)]$, but the trace of
$\tilde{H}_{I}(t_1-t_2)$ vanishes, 
i.e., ${\rm Tr}_{R}[\tilde{H}_{I}(t_1-t_2)R_0]=0$,  
because of Eq.~(\ref{traceless_1st_order}). 
Thus Eq.~(\ref{non-Markovian_1st_order}) vanishes in the Markovian limit.
But the situation differs for a
non-Markovian environment as the environment operator in
 $\tilde{H}_{I}(t_1-t_2)$  may, in general, be correlated with that in 
$U(t_2,0)$.
Therefore, the evolution from  $\rho_T(0)$ to $\rho_T(t_2)$ 
under the influence of interaction Hamiltonian 
in the presence of the reservoir needs to be taken into account 
before the trace over the environment is performed in 
Eq.~(\ref{non-Markovian_1st_order}). 
We emphasize here that it is this nonlocal environment (bath) memory term,
Eq.~(\ref{non-Markovian_1st_order}), that vanishes in the Markovian
case but makes the evolution equation of the two-time  
CF's of the system operators deviate from the QRT. 
As we aim to obtain an evolution equation of the two-time CF's of the
system operators, valid up to second order in the interaction
Hamiltonian, we need to find $\rho_T(t_2)$ only up to 
first order in the interaction Hamiltonian.   
So substituting 
$\rho_{T}(t_2)= e^{-iH_{0}t_2/\hbar }\tilde{\rho}_{T}(t_2)e^{iH_{0}t_2/\hbar }$
with the expression
\begin{equation}
\tilde{\rho}_{T}\left( t_2\right) \approx \tilde{\rho}_{T}\left( 0\right) 
-\frac{i}{\hbar }\int_{0}^{t_2}d\tau\left[ \tilde{H}_{I}\left(
\tau\right) ,\tilde{\rho}_{T}\left( t_2\right) \right] 
\label{rhoT_t2}
\end{equation}%
for $\tilde{\chi}_T(0)=\chi_T(0)=B\rho_T(t_2)$ 
in Eq.~(\ref{non-Markovian_1st_order}),
we then obtain up to second order in the interaction Hamiltonian (in
the system-environment coupling strength)     
\begin{eqnarray}
&&-\frac{1}{\hbar^2}\int_0^{t_2}d\tau\, {\rm Tr}_{S\otimes R}\left(Ae^{-iH_{0}t/\hbar}\, 
\left[ \tilde{H}_{I}\left(
t\right), \right. \right. \nonumber\\
&&\quad \quad\left. \left. B\,e^{-iH_{0}t_2/\hbar} [\tilde{H}_{I}(\tau),\tilde{\rho}_{T}(t_2)]e^{iH_{0}t_2/\hbar} \right]\, e^{iH_{0}t/\hbar }\right),
\nonumber \\
&=& -\frac{1}{\hbar^2}\int_0^{t_2}d\tau\, 
{\rm Tr}_{S\otimes R}\left( A
\left[ H_{I},\right. \right. \nonumber\\
&&\quad \quad\left. \left. e^{-iH_{0}t/\hbar}\,
B[\tilde{H}_{I}(\tau-t_2),\rho_{T}(t_2)] \, e^{iH_{0}t/\hbar }\right]\right),
\label{extra1}
\end{eqnarray}
where the first order term in the interaction Hamiltonian coming from $\tilde{\rho}_T(0)$ term in Eq.~(\ref{rhoT_t2}) 
has been dropped because of Eq.~(\ref{traceless_1st_order}). 
Since the product of two $H_I$ already appear explicitly in Eq.~(\ref{extra1}),
we may then transform Eq.~(\ref{extra1}) into Heisenberg
representation with the evolution equation $\exp(iHt/\hbar)\approx
\exp(iH_0t/\hbar)+{\cal O}(H_I)$. 
Furthermore, by writing out the commutators explicitly and rearranging the
Heisenberg operator terms,   
the resultant equation from Eq.~(\ref{extra1}) then becomes
\begin{eqnarray}
&& -\frac{1}{\hbar^2}\int_0^{t_2}d\tau\, 
{\rm Tr}_{S\otimes R}\left( 
\{\tilde{H}_I(\tau-t_1)[H_I,A]\}(t_1)B(t_2)\rho_{T}(0) \right. \nonumber\\
&&\quad\quad\quad
\left. +\{[A,H_I]\}(t_1)\{B\tilde{H}_I(\tau-t_2)\}(t_2)\rho_T(0) \right).
\label{extra2}
\end{eqnarray}
The first term in Eq.~(\ref{extra2}) is ready to combine with the second term in Eq.~(\ref{2time_evol_1_3}) to extend the integration from $0$ to $t_1$. 
Similarly, one may rewrite the last term in Eq.(\ref{extra2}) using the relation $B\tilde{H}_I(\tau-t_2)=\tilde{H}_I(\tau-t_2)B+[B,\tilde{H}_I(\tau-t_2)]$ so that the first new term can be combined with last term in Eq.~(\ref{2time_evol_1_3}) to extend the integration from $0$ to $t_1$.
        
Putting all the resultant terms together, we obtain the evolution equation of the two-time CF's valid to second order in the interaction Hamiltonian as
\begin{eqnarray}
&&{d\left\langle A(t_{1}) B(t_2)\right\rangle }/{dt_{1}} \nonumber \\
&=&({i}/{\hbar}){\rm Tr}_{S\otimes R}\left( \{[{H}_S, {A}] \}(t_1) 
B(t_2)\rho_{T}(0)\right) \nonumber \\
&&+\frac{1}{\hbar^2 }\int_{0}^{t_1}d\tau {\rm Tr}_{S\otimes R} \nonumber\\
&&\quad
\left( \{\tilde{H}_I(\tau-t_1)[A,H_I]\}(t_1)B(t_2)\rho_{T}(0) \right. \nonumber\\
&&\quad
+\left.\{[H_I,A]\tilde{H}_I(\tau-t_1)\}(t_1)B(t_2)\rho_{T}(0) \right)
\nonumber\\
&&
+\frac{1}{\hbar^2 }\int_{0}^{t_2}d\tau {\rm Tr}_{S\otimes R} \nonumber\\
&&\quad \left(\{[H_I,A]\}(t_1)\{[B,\tilde{H}_I(\tau-t_2)]\}(t_2)\rho_T(0)\right).
\label{2time_evol_eq}
\end{eqnarray}
Compared to Eq.~(\ref{1time_evol_eq}), it is the existence of the last
term in Eq.~(\ref{2time_evol_eq}) that makes the QRT invalid. 
Equation (\ref{2time_evol_eq}) is the main result of this paper. 
The derivation is based on perturbative quantum master
equation approach, so non-Markovian open quantum system models that are
not exactly solvable can use our derived evolution equation to 
obtain the time evolutions of 
their two-time CF's of system operators, valid to second order in
the system-environment interaction. 
In the derivation of Eqs.~(\ref{1time_evol_eq}) and (\ref{2time_evol_eq}),
we have also used the assumption of a factorized initial system-bath state
$\rho_{T}(0)=\tilde{\rho}_{T}(0)=\rho(0)\otimes R_0$ and the
condition of 
${\rm Tr}_R[\tilde{H}_{I}(t) R_0]=0$, Eq.~(\ref{traceless_1st_order}),
to eliminate the first-order term.
Since the form and nature of the Hamiltonians are not specified,
Eq.~(\ref{2time_evol_eq}) can be used to calculate the two-time CF's 
for non-Markovian open quantum systems with 
multi-level discrete or continuous Hilbert
spaces, interacting with bosonic and/or
fermionic environments.  
The procedure and the degrees of difficulty 
to apply Eq.~(\ref{2time_evol_eq}) to a open quantum system model
(by  taking into account 
nonlocal bath memory effects and 
tracing out the bath degrees of freedom for 
factorized system-bath initial states) to obtain the two-time
CF's of system operator 
are similar to those for the evaluation of the reduced density matrix
of a second-order time-convolutionless non-Markovian 
quantum master equation [e.g., Eq.~(\ref{time_convolutionless_ME})].
We will explicitly apply the evolution equation (\ref{2time_evol_eq})
to a general model of a quantum system coupled to a
finite-temperature bosonic environment in Sec.~\ref{sec:bosonic}
and a specific model of two-level system in Sec.~\ref{sec:spin-boson}.
Open quantum systems coupled to fermionic reservoirs
(environments) could, for example,
be quantum
dots or other nanostructure systems coupled (connected) to
nonequilibrium electron reservoirs (electrodes or leads) in the
electron transport problems \cite{Kleinekathofer04,Welack06,Harbola06,
  Goan01, Li04, Utami04, Zedler09,Gudmundsson09,Jin10}. 
The evolution equation
(\ref{2time_evol_eq}) can also be used to calculate the non-Markovian 
two-time CF's in such systems.
In summary, our evolution equation (\ref{2time_evol_eq}) can be
applied to a wide range 
of system-environment models with 
any factorized (separable) system-environment initial states (pure or mixed).

\section{Evolution equations for thermal bosonic bath models}
\label{sec:bosonic}
To make contact with Refs.~\cite{Alonso05,Vega06,Alonso07}, we 
consider a quantum system coupled to a
bosonic environment with a general Hamiltonian of the form
\begin{eqnarray}
H&=&H_{S}+\sum_{\lambda }\hbar g_{\lambda }\left( L^{\dagger} a_{\lambda
}+La_{\lambda }^{\dagger}\right) +\sum_{\lambda }\hbar\omega_{\lambda
}a_{\lambda }^{\dagger}a_{\lambda }, 
\label{Hamiltonian_L}
\end{eqnarray}
where the system coupling operator $L$ acts on the Hilbert space of
the system, 
$a_{\lambda }$ and $a_{\lambda }^{\dagger}$ are the annihilation and creation
operators on the bosonic environment Hilbert space, and $g_{\lambda }$
and $\omega _{\lambda }$ are respectively the coupling strength and the
frequency of the $\lambda$th environmental oscillator.

Applying Eq.~(\ref{Hamiltonian_L}) to Eqs.~(\ref{1time_evol_eq}) and
(\ref{2time_evol_eq}) and after tracing over the environmental degrees
of freedom for factorized (separable) system-bath initial states, 
we arrive at the second-order evolution equations of the single-time expectation values
\begin{eqnarray}
&&
{d\left\langle A\left( t_{1}\right) \right\rangle }/{dt_{1}} 
\nonumber \\
&=&
({i}/{\hbar }){\rm Tr}_{S}\left( \{[H_{S},A]\}(t_1) \rho(0) \right)
\nonumber \\
&&+
\int_{0}^{t_{1}}d\tau {\rm Tr}_S
\nonumber \\
&&\quad \left( \alpha^{\ast }( t_{1}-\tau)
\{\tilde{L}^{\dagger}(\tau -t_{1})[{A},{L}]\}( t_{1}){\rho}(0) \right.
\nonumber \\
&&\quad
+\alpha( t_{1}-\tau )\{ [ {L}^{\dagger},A] \tilde{L}(\tau -t_{1})\}(t_{1}) 
\rho( 0) 
  \nonumber \\
&&\quad+
\beta^{\ast }(t_{1}-\tau)\{\tilde{L}(\tau -t_{1})[A,L^{\dagger}]\}(t_{1})
\rho(0) 
\nonumber \\
&&
\quad+
\left.
\beta(t_{1}-\tau)\{[L,A] \tilde{L}^{\dagger}(\tau -t_{1})\}(t_{1}) 
\rho(0) \right), 
\label{1time_evol_eq_f}
\end{eqnarray}
and of the two-time CF's
\begin{eqnarray}
&&{d\left\langle A\left( t_{1}\right) B\left( t_{2}\right)
\right\rangle }/{dt_{1}} \nonumber \\ 
&=&
({i}/{\hbar}){\rm Tr}_{S}\left( \{[{H}_{S}, A]\}( t_{1}){B}( t_{2}) 
\rho(0) \right)  \nonumber \\
&&+
\int_{0}^{t_{1}}d\tau {\rm Tr}_S
\nonumber \\
&&\quad\left( \alpha^{\ast }( t_{1}-\tau)
\{\tilde{L}^{\dagger}(\tau -t_{1})[{A},{L}]\}( t_{1}) 
{B}(t_{2}){\rho}(0) \right.
\nonumber \\
&&\quad+
\alpha( t_{1}-\tau )\{ [ {L}^{\dagger},A] \tilde{L}(\tau -t_{1})\} ( t_{1}) B( t_{2}) 
\rho( 0) 
\nonumber \\
&&\quad+
\beta^{\ast }(t_{1}-\tau)\{\tilde{L}(\tau -t_{1})[A,L^{\dagger}]\}(t_{1}) B(t_{2}) 
\rho(0) 
\nonumber \\
&&\quad+
\left. 
\beta(t_{1}-\tau)\{[L,A] \tilde{L}^{\dagger}(\tau -t_{1})\}(t_{1}) 
B(t_{2})\rho(0) 
\right)  
\nonumber \\
&&+
\int_{0}^{t_{2}}d\tau 
{\rm Tr}_S \nonumber \\
&&\quad\left( \alpha(t_{1}-\tau) \{ [ L^{\dagger},A]\}(t_{1}) 
\{ [B,\tilde{L}(\tau -t_{2})]\}(t_{2}) \rho(0) \right.
\nonumber \\
&&\hspace{-0.3cm}+
\left. \beta( t_{1}-\tau)
\{ [L,A]\}(t_{1})\{[B,\tilde{L}^{\dagger}(\tau -t_{2})]\} (t_{2}) 
\rho(0) \right). 
\label{2time_evol_eq_f}
\end{eqnarray}
Here $\tilde{L}(t)=\exp(iH_St/\hbar)L\exp(-iH_St/\hbar)$ is the system operator in the interaction picture with respect to $H_S$, and 
\begin{eqnarray}
\alpha \left( t_{1}-\tau \right) &=&\sum_{\lambda }\left( \bar{n}_{\lambda }+1\right)
\left\vert g_{\lambda }\right\vert ^{2}
e^{-i\omega _{\lambda }\left( t_{1}-\tau \right)},
\label{CFalpha}\\ 
\beta \left( t_{1}-\tau \right)
&=&\sum_{\lambda }\bar{n}_{\lambda }\left\vert g_{\lambda }\right\vert ^{2}e^{i\omega
_{\lambda }\left( t_{1}-\tau \right) }
\label{CFbeta}
\end{eqnarray}
are the environment CF's with $\alpha(t_1-\tau)=\left\langle
  \sum_{\lambda}g_{\lambda}\tilde{a}_{\lambda}(t_1)\sum_{\lambda'}g_{\lambda'}\tilde{a}^{\dagger}_{\lambda'}(\tau)\right\rangle_R$
and $\beta(t_1-\tau)=\left\langle
  \sum_{\lambda}g_{\lambda}\tilde{a}^{\dagger}_{\lambda}(t_1)\sum_{\lambda'}g_{\lambda'}\tilde{a}_{\lambda'}(\tau)\right\rangle_R$,
where $\tilde{a}_{\lambda}(t_1)=a_{\lambda}e^{-i\omega_\lambda t_1}$
and
$\tilde{a}^{\dagger}_{\lambda}(t_1)=a^{\dagger}_{\lambda}e^{i\omega_\lambda
  t_1}$ are the environment operators in the interaction
picture, and the symbol $\langle \cdots \rangle_R$ denotes taking a trace with
respect to the density matrix of 
the thermal bosonic reservoir (environment). 
The thermal mean occupation number $\bar{n}_\lambda$ of the
bosonic environment
oscillators in Eqs.~(\ref{CFalpha}) and (\ref{CFbeta}) is
$\bar{n}_\lambda=(e^{\hbar\omega_\lambda/k_BT}-1)^{-1}$. 

The evolution equations (\ref{1time_evol_eq_f}) and (\ref{2time_evol_eq_f})
for a non-Markovian bosonic environment have been presented in
Ref.~\cite{Goan10} without any derivation. In this paper, the detailed
derivation of the evolution equations is given.
Furthermore, the two-time CF evolution equation, Eq.~(\ref{2time_evol_eq}),
applicable for both bosonic and fermionic environments and applicable
for more general form of system-environment interaction Hamiltonian 
has not been published in the literature yet.

As mentioned, the two-time evolution equations derived 
in Refs.~\cite{Alonso05,Vega06,Alonso07} is, strictly speaking,
applicable only for a
zero-temperature environment. However, these equations
were used to calculate the
two-time CF's of system observables of dissipative spin-boson models
at finite temperatures.    
This is possible only for the dissipative spin-boson models 
with Hermitian system coupling
operators, $L=L^\dagger$. We will discuss this point in details
in subsection \ref{sec:comparison}.
In contrast, our bosonic evolution equations,
Eqs.~(\ref{1time_evol_eq_f}) and (\ref{2time_evol_eq_f}), are valid
for both a Hermitian and a non-Hermitian system coupling operators and
can be used to calculate the two-time CF's for any factorized (separable)
system-reservoir initial state and for any arbitrary temperature as
long as the assumption of the weak system-environment coupling still
holds. 

In Ref.~\cite{Goan10}, we used Eqs.~(\ref{1time_evol_eq_f}) and 
(\ref{2time_evol_eq_f}) to calculate
the finite-temperature 
single-time expectation values and two-time CF's  for a non-Markovian
pure-dephasing spin-boson model of 
\begin{equation}
H_S=(\hbar\omega_S/2)\sigma_z, \quad\quad
L=\sigma_z=L^{\dagger}.  
\label{SpinBoson}  
\end{equation}
Since the non-Markovian dynamics of this exactly solvable
pure-dephasing model can be cast into a time-local,
convolutionless form and $[L,H_S]=0$, the results obtained by
our second-order evolution equations turn out to be
exactly the same as the exact results obtained by the direct
operator evaluation. 
However, these results significantly differ from the non-Markovian
two-time CF's obtained by wrongly directly applying the quantum
regression theorem (QRT). 
This demonstrates the validity of
the evolution equations (\ref{1time_evol_eq_f}) and 
(\ref{2time_evol_eq_f}).
But the system coupling operators $L$ of this pure dephasing model
\cite{Goan10} and the examples calculated in
Refs.~\cite{Alonso05,Vega06,Alonso07} are all Hermitian, i.e., $L^{\dagger}=L$.   
So we will present in Sec.~\ref{sec:spin-boson} the calculations of
one-time averages and two-time CF's for a thermal spin-boson model
with $L\neq L^\dagger$, for which only the evolution equations
(\ref{1time_evol_eq_f}) and (\ref{2time_evol_eq_f}), rather than those
in Refs.~\cite{Alonso05,Vega06,Alonso07}, are applicable.

\subsection{Comparison and discussion}  
\label{sec:comparison}
We discuss in the following the connection of our derived two-time 
evolution equation (\ref{2time_evol_eq_f}) with those presented in 
Refs.~\cite{Alonso05,Vega06,Alonso07}.
In Ref.~\cite{Vega06}, a master equation conditioned on initial coherent states of the environment in Bargmann representation, $(z_0,z'_0)$, was  derived in the weak system-environment coupling limit. Provided that the whole set of the initial conditions of the system of interest, $|\psi(z^*_0)\rangle$, and the statistical probability ${\cal J}(z_0,z^*_0)$ for the member $|\psi(z^*_0)\rangle|z_0\rangle$ of the statistical ensemble are known, this master equation 
with $z'_0=z_0$ is capable of evaluating the evolution of 
{\em single-time} expectation values for general initial conditions, including initially correlated mixed states between the system and environment.  
However, the evolution equations of {\em two-time (multi-time)} CF's of system observables, Eq.~(6) in Ref.~\cite{Alonso05}, Eq.~(31) in Ref.~\cite{Vega06} and Eq.~(60) in Ref.~\cite{Alonso07}, were derived for an initial vacuum state of the environment and an initial pure state of the system of interest.  
As a result, these two-time evolution equations are valid only for a zero-temperature environment 
(if the system coupling operator is not Hermitian, i.e., $L\neq L^\dagger$; see discussions below). 
Compared with the corresponding zero-temperature two-time (multi-time) evolution equations 
derived in Refs.~\cite{Alonso05,Vega06,Alonso07}, our finite-temperature 
two-time evolution equation 
(\ref{2time_evol_eq_f}) is valid for any initial factorized (separable) states (pure or mixed) at finite temperatures and for both Hermitian and non-Hermitian system coupling operators. The extra terms containing the bath CF $\beta(t_1-\tau)$ or $\beta^*(t_1-\tau)$ are due to the finite-temperature thermal environment. If we take the zero-temperature limit by letting  $\bar{n}_\lambda=0$ and thus $\beta(t_1-\tau)=\beta^*(t_1-\tau)=0$, as well as consider the initial pure-state case by letting ${\rm Tr}_S[\cdots\rho(0)]\to \langle \psi(0)|\cdots |\psi(0)\rangle$, then Eqs.(\ref{1time_evol_eq_f}) and (\ref{2time_evol_eq_f}) reduce exactly to their corresponding zero-temperature pure-state evolution equations in Refs.~\cite{Alonso05,Vega06,Alonso07}.

However, calculations of the two-time CF's of system observables of
dissipative spin-boson models in finite-temperature thermal baths (rather than zero-temperature baths) were presented in Refs.~\cite{Alonso05,Vega06,Alonso07} 
even though in their derivations of the two-time (multi-time) evolution equations, the bath CF is given in its zero-temperature form, 
\begin{equation}
\alpha_0(t-\tau)=\sum_{\lambda }
\left\vert g_{\lambda }\right\vert ^{2}
e^{-i\omega _{\lambda }\left( t_{1}-\tau \right)}. 
\label{CFalpha0}
\end{equation}
This is only possible due to the reason that the system coupling operator is Hermitian, $L=L^\dagger$, in the thermal bath examples presented in Refs.~\cite{Alonso05,Vega06,Alonso07}. 
One may understand this as follows.   
It is known that the finite-temperature density matrix operator of a thermal bath can be canonically mapped onto the zero-temperature density operator (the vacuum) of a larger (hypothetical) environment \cite{Semenoff83,Yu04}. 
By mapping the total Hamiltonian Eq.~(\ref{Hamiltonian_L})  and an
initial thermal state to an extended total system with a vacuum state,
the finite-temperature problem can be reduced to a zero-temperature
problem, and the resultant pure state $\psi_t=|\psi_t(z^*,w^*)\rangle$
for the system of interest  satisfies the following linear
finite-temperature non-Markovian stochastic Schr\"odinger equation
with two independent noises $z^*_t$ and $w^*_t$ \cite{Diosi98,Yu04}:
\begin{eqnarray}  
\frac{\partial}{\partial t}\psi_t&=&-iH_S\psi_t+L z^*_t\psi_t-L^\dagger\int_0^t d\tau \alpha(t-\tau)\frac{\delta\psi_t}{\delta z^*_\tau} \nonumber \\
&&+L^\dagger w^*_t\psi_t-L\int_0^t d\tau \beta(t-\tau)\frac{\delta\psi_t}{\delta w^*_\tau}, 
\label{SSEfiniteT}
\end{eqnarray}
where 
\begin{eqnarray} 
z^*_t&=&-i\sum_\lambda \sqrt{\bar{n}_{\lambda }+1}\, g^*_\lambda z^*_\lambda e^{i\omega_\lambda t},
\label{zt}\\
w^*_t&=&-i\sum_\lambda \sqrt{\bar{n}_{\lambda }}\, g^*_\lambda w^*_\lambda e^{-i\omega_\lambda t}
\label{wt}
\end{eqnarray}
are two independent, colored, complex Gaussian noises with zero mean 
and satisfy 
\begin{eqnarray}
{\cal M}[z_tz_\tau]={\cal M}[z^*_tz^*_\tau]=0,&&\quad {\cal M}[z^*_tz_\tau]=\alpha(t-\tau);
\\
{\cal M}[w_tw_\tau]={\cal M}[w^*_tw^*_\tau]=0,&&\quad {\cal M}[w^*_tw_\tau]=\beta(t-\tau).
\end{eqnarray}
Here,  $z^*_\lambda$  and $w^*_\lambda$ are coherent state variables of the extended environment in Bargmann representation, ${\cal M}[\cdots]$ denotes the statistical average over the Gaussian processes $z^*_t$ and $w^*_t$, and the bath CF's $\alpha(t-\tau)$ and $\beta(t-\tau)$ are defined in Eqs.~(\ref{CFalpha}) and (\ref{CFbeta}), respectively.
In the zero-temperature limit $T\to 0$, the mean thermal occupation number of quanta in mode $\lambda$ approaches zero, i.e., $\bar{n}_{\lambda } \to 0$. 
and thus $\beta(t-\tau)\to 0$ and $\alpha(t-\tau)\to \alpha_0(t-\tau)$ 
that is defined in Eq.~(\ref{CFalpha0}). 
In this case, the noises, Eqs.~(\ref{zt}) and (\ref{wt}), become
 $z^*_t=-i\sum_\lambda g^*_\lambda z^*_\lambda e^{i\omega_\lambda t}$
 and  $w^*_t=0$, and  
the finite-temperature equation (\ref{SSEfiniteT}) reduces to the
simple zero-temperature equation \cite{Diosi98,Yu04} 
\begin{equation}
\frac{\partial}{\partial t}\psi_t=-iH_S\psi_t+L z^*_t\psi_t-L^\dagger\int_0^t d\tau \alpha_0(t-\tau)\frac{\delta\psi_t}{\delta z^*_\tau}.
\label{SSEzeroT}
\end{equation}

Now consider the case of a Hermitian system coupling operator $L=L^\dagger$. 
The finite-temperature equation (\ref{SSEfiniteT}) can be simplified considerably by introducing the sum process $u^*_t=z^*_t+w^*_t$ that has a zero mean and satisfies
\begin{eqnarray}
{\cal M}[u_tu_\tau]&=&{\cal M}[u^*_tu^*_\tau]=0;\\
{\cal M}[u^*_tu_\tau]&=&\alpha_{\rm eff}(t-\tau) \nonumber\\
&=&\alpha(t-\tau)+\beta(t-\tau) \nonumber \\
&=& \sum_\lambda |g_\lambda|^2
\left\{\coth\left({\hbar\omega_\lambda}/{2k_{B}T}\right)\cos[\omega_\lambda(t-\tau)]\right.
\nonumber\\
&& \hspace{1.5cm}\left.-i\sin[\omega_\lambda(t-\tau)]\right\}. 
\label{CFalpha_eff}  
\end{eqnarray}
In terms of the single noise process $u^*_t$, the linear finite-temperature non-Markovian stochastic Schr\"odinger equation (\ref{SSEfiniteT}) for the case of a Hermitian coupling operator $L=L^\dagger$ takes the simple form of the zero-temperature equation (\ref{SSEzeroT}) with the replacement of $z^*_t$ by $u^*_t$ and 
$\alpha_0(t-\tau)$ by $\alpha_{\rm eff}(t-\tau)=\alpha(t-\tau)+\beta(t-\tau)$ 
defined in Eq.~(\ref{CFalpha_eff}). 
It is for this reason of the Hermitian coupling operator
$L=L^\dagger=\sigma_x$ in the dissipative spin-boson model with a
thermal environment that the two-time CF's of the system observables
can be evaluated with the evolution equations derived in
Refs.~\cite{Alonso05,Vega06,Alonso07}. 
In other words, if the system operator coupled to the environment is
not Hermitian $L\neq L^\dagger$,  
the two-time (multi-time) differential evolution equations presented in 
Refs.~\cite{Alonso05,Vega06,Alonso07} are valid only for a zero-temperature
environment.

In contrast, our two-time evolution equation (\ref{2time_evol_eq_f}) is valid for more general finite-temperature cases where the system coupling operator is not a Hermitian operator, i.e., $L\neq L^\dagger$. 
In the case of $L\neq L^\dagger$, 
our two-time evolution equation contains additional finite-temperature $\beta(t_1-\tau)$ and $\beta^*(t_1-\tau)$ terms which can not be combined and simplified to a simpler form as the zero-temperature evolution equation derived in Refs.~\cite{Alonso05,Vega06,Alonso07}.
For a Hermitian coupling operator $L=L^{\dagger}$, one can see that besides the replacement of a more general system state with an initial pure system state by letting $\langle \psi(0)|\cdots |\psi(0)\rangle \to {\rm Tr}_S[\cdots\rho(0)]$, the finite-temperature evolution equation (\ref{2time_evol_eq_f}) reduces to its zero-temperature counterpart in Refs.~\cite{Alonso05,Vega06,Alonso07} with the effective bath CF given by  
$\alpha_{\rm eff}=\alpha(t_1-\tau)+\beta(t_1-\tau)$ defined in Eq.~(\ref{CFalpha_eff}). This demonstrates explicitly why the zero-temperature 
two-time evolution equations derived in Refs.~\cite{Alonso05,Vega06,Alonso07} can be used to calculate the system operator CF's for a thermal spin-boson model with a Hermitian system coupling operator.

\subsection{Conditions for the QRT to hold} 
\label{sec:QRT_hold}

As mentioned in subsection \ref{sec:QRT},
QRT is a very useful procedure that enables one
(in certain circumstances) to calculate for two-time
(multi-time) CF's of system operators from the knowledge of the
evolution equations of 
single-time expectation values. 
One can notice that if the last two terms in
Eq.~(\ref{2time_evol_eq_f}) vanish, then the non-Markovian
single-time and two-time evolution equations (\ref{1time_evol_eq_f}) and (\ref{2time_evol_eq_f}) will have the same form with the same evolution coefficients and thus the QRT can apply.  As expected, these two terms vanish in the Markovian case since the time integration of the corresponding $\delta$-correlated reservoir CF's, 
$\alpha(t_1-\tau)\propto \delta(t_1-\tau)$ and 
$\beta(t_1-\tau)\propto \delta(t_1-\tau)$, over the variable $\tau$ in the domain from $0$ to $t_2$ is zero as $t_1>t_2$.
So from Eq.~(\ref{2time_evol_eq_f}), in the weak system-environment
coupling case the QRT holds when (i) $[ L^{\dagger},A]=0$ or
$[B,\tilde{L}(\tau -t_{2})]=0$ at the zero temperature, (ii) at finite
temperatures, in addition to condition (i), the following condition
also needs to be satisfied: $[L,A]=0$ or $[B,\tilde{L}^{\dagger}(\tau -t_{2})]=0$, (iii) in the Markovian case where the bath CF's are 
$\delta$-correlated in time.   

Note that in some models, certain CF's, which formally obey the QRT but with evolution equations coupled with those of other CF's that do not obey the QRT,
may yield solutions different from those given by the QRT \cite{Vega06}.

\section{Application to a thermal spin boson model with $L\neq L^\dagger$}
\label{sec:spin-boson}
To illustrate the usage of the equations we have derived, we apply them to the problem of a two-level system coupled to a thermal reservoir, in which  $L\neq L^{\dagger}$. 
We consider Hamiltonian, Eq.~(\ref{Hamiltonian_L}), with $H_S=(\hbar\omega_A/2)\sigma_z$, a coupling operator $L=\sigma_-$ and a system-environment interaction Hamiltonian whose magnitude is small enough to be considered as a perturbation. Since the coupling operator $L\neq L^{\dagger}$ is not Hermitian, 
the two-time (multi-time) evolution equations derived in Refs.~\cite{Alonso05,Vega06,Alonso07} are not applicable and 
our evolution equation (\ref{2time_evol_eq_f}) should be employed to obtain the time evolutions of the two-time CF's. 

\subsection{Single-time expectation values}

Before calculating the two-time CF's, it is instructive to
obtain the master equation of the reduced system density matrix for
the model. 
Transferring from the interaction picture back to the Schr\"odinger
picture for Eq.~(\ref{time_convolutionless_ME})   
and using the general Hamiltonian, Eq.~(\ref{Hamiltonian_L}), we obtain 
the second-order time-convolutionless non-Markovian master equation
for the reduced density matrix $\rho(t)$ as
\begin{eqnarray}
\frac{d \rho(t)}{dt} 
&=&\frac{1}{i\hbar}\left[H_S,\rho(t)\right] 
\nonumber \\
&&
\hspace{-1cm}
-\int_0^t d\tau \{\alpha(t-\tau)[L^\dagger\tilde{L}(\tau-t)\rho(t)
-\tilde{L}(\tau-t)\rho(t)L^\dagger] \nonumber\\
&&
\hspace{-0.5cm}
 +\beta(t-\tau)[L\tilde{L}^\dagger(\tau-t)\rho(t)
-\tilde{L}^\dagger(\tau-t)\rho(t)L] \nonumber \\
&&\hspace{-0.5cm} +{\rm h.c.} \},
\label{master_eq}
\end{eqnarray}
where $\alpha(t-\tau)$ and $\beta(t-\tau)$ are defined in
Eqs.~(\ref{CFalpha}) and (\ref{CFbeta}) respectively, 
h.c. indicates the hermitian conjugate of previous terms, and an operator with a tilde on the top indicates that it is an operator in the interaction picture.
The only real assumption used to obtain Eq.~(\ref{master_eq})
valid to second order in the system-environment coupling strength 
is the total density matrix
factorizing at the initial time $t=0$, $\rho_T(0)=\rho(0)\otimes R_0$.
Taking $L=\sigma _{-}$, $L^{\dagger }=\sigma _{+}$, then
$\tilde{L}(t)=\sigma _{-}e^{-i\omega _{A}t}$, and
$\tilde{L}^{\dagger}(t)=\sigma _{+}e^{i\omega _{A}t}$, and
substituting them into Eq.~(\ref{master_eq}), we obtain  
\begin{eqnarray}
\frac{d\rho(t)}{dt}&=&-i\frac{\omega_A}{2}\left[\sigma_z,\rho(t)\right]\nonumber\\
&&
\hspace{-1cm}
-\left\{\Gamma _{1}(t)\left(\sigma_+\sigma_
-\rho(t)-\sigma_-\rho(t)\sigma_+\right)\right.
\nonumber\\
&&
\hspace{-1cm}
\left. +\Gamma _{2}(t)\left(\sigma_-\sigma_+\rho(t)
-\sigma_+\rho(t)\sigma_-\right)
+h.c.\right\},
\label{master_eq_Schrodinger}
\end{eqnarray} 
where
\begin{eqnarray}
\Gamma _{1}(t) &=&\int_{0}^{t}d\tau \alpha (t-\tau)e^{+i\omega _{A}\left(t-\tau \right) },  
\label{Gamma_1} \\
\Gamma _{2}(t) &=&\int_{0}^{t}d\tau \beta (t-\tau)e^{-i\omega _{A}\left(t-\tau \right) }.  
\label{Gamma_2}
\end{eqnarray}
The master equation (\ref{master_eq_Schrodinger}) 
is a time-local and convolutionless differential equation.
The effect of the non-Markovian environment 
on the second-order master equation (\ref{master_eq_Schrodinger})
is taken into account by the time-dependent coefficients $\Gamma _{1}(t_{1})$ and $\Gamma _{2}(t_{1})$ defined in Eqs.~(\ref{Gamma_1}) and (\ref{Gamma_2}) 
instead of memory integrals. 
Likewise, the evolution equations of single-time expectation values and two-time CF's of system operators valid to the second order 
are also expected to be convolutionless.
Taking again $L=\sigma _{-}$, $L^{\dagger }=\sigma _{+}$, then
$\tilde{L}(t)=\sigma _{-}e^{-i\omega _{A}t}$, and
$\tilde{L}^{\dagger}(t)=\sigma _{+}e^{i\omega _{A}t}$, and using the
commutation relation between the Pauli matrices, 
we obtain  
straightforwardly from Eq.~(\ref{1time_evol_eq_f}) the following evolution
equations of the single-time expectation values of system operators as 
\begin{eqnarray}
d\langle \sigma _{+}(t_{1})\rangle /dt_{1} &=&i\omega _{A}\langle \sigma
_{+}(t_{1})\rangle  \nonumber \\
&&-\left[ \Gamma _{1}^{\ast }(t_{1})+\Gamma _{2}(t_{1})\right] \langle
\sigma _{+}(t_{1})\rangle,  \label{one_evol_+} \\
d\langle \sigma _{-}(t_{1})\rangle /dt_{1} &=&-i\omega _{A}\langle \sigma
_{-}(t_{1})\rangle  \nonumber \\
&&-\left[ \Gamma _{1}(t_{1})+\Gamma _{2}^{\ast }(t_{1})\right] \langle
\sigma _{-}(t_{1})\rangle,  \label{one_evol_-} \\
d\langle \sigma _{z}(t_{1})\rangle /dt_{1} &=&-\left[ \Gamma
_{1}(t_{1})+\Gamma _{1}^{\ast }(t_{1})+\Gamma _{2}^{\ast }(t_{1})+\Gamma
_{2}(t_{1})\right] \nonumber \\
&&\quad\quad \times \langle \sigma _{z}(t_{1})\rangle  \nonumber \\
&&-\left[ \Gamma _{1}(t_{1})+\Gamma _{1}^{\ast }(t_{1})-\Gamma _{2}^{\ast
}(t_{1})-\Gamma _{2}(t_{1})\right].   \nonumber \\
\label{one_evol_z}
\end{eqnarray}
Equations (\ref{one_evol_+})-(\ref{one_evol_z}) 
can also be obtained directly from the master equation
(\ref{master_eq_Schrodinger}) through ${d\left\langle A\left( t_{1}\right) \right\rangle }/{dt_{1}}={\rm Tr}_{S}\left[ A({d\rho(t_1)}/{dt_1})\right]$.

\subsection{Two-time CF's}
For the evaluations of the two-time CF's of the system observables, we consider the following four cases.

{\em {Case 1:}} $[L^{\dagger },A]=0$ or $[B,\tilde{L}(t)]=0$; and $[L,A]=0$
or $[B,\tilde{L}^{\dagger }(t)]=0$. In this case, let $A=\sigma _{i}$, $%
B=\sigma _{i}$ with $i=+,-$. 
Applying the commutation relations between the Pauli matrices and the
definition of ${\rm Tr}_S[\sigma_i(t_1) \sigma_i(t_2)\rho(0)]=\langle
\sigma_i(t_1) \sigma_i(t_2) \rangle$ to the right-hand side of the
two-time evolution equation (\ref{2time_evol_eq_f}), 
we then obtain
\begin{eqnarray}
d\langle \sigma _{+}(t_{1})\sigma _{+}(t_{2})\rangle /dt_{1} &=&i\omega
_{A}\langle \sigma _{+}(t_{1})\sigma _{+}(t_{2})\rangle  \nonumber \\
&&-\left[ \Gamma _{1}^{\ast }(t_{1})+\Gamma _{2}(t_{1})\right] \langle
\sigma _{+}(t_{1})\sigma _{+}(t_{2})\rangle ,  \nonumber \\
 \label{2time_evol_++} \\
d\langle \sigma _{-}(t_{1})\sigma _{-}(t_{2})\rangle /dt_{1} &=&-i\omega
_{A}\langle \sigma _{-}(t_{1})\sigma _{-}(t_{2})\rangle  \nonumber \\
&&-\left[ \Gamma _{1}(t_{1})+\Gamma _{2}^{\ast }(t_{1})\right] \langle
\sigma _{-}(t_{1})\sigma _{-}(t_{2})\rangle ,  \nonumber \\
\label{2time_evol_--}
\end{eqnarray}%
where  
$\Gamma _{1}(t_{1})$ and $\Gamma _{2}(t_{1})$ are defined in Eqs.~(\ref%
{Gamma_1}) and (\ref{Gamma_2}), respectively. One can see that the evolution
equations of the two-time CF's, Eqs.~(\ref{2time_evol_++}) and (\ref%
{2time_evol_--}), have the same forms as the evolution equations of the
single-time expectation values $\langle \sigma _{+}(t_{1})\rangle $ and $%
\langle \sigma _{-}(t_{1})\rangle $, Eqs.~(\ref{one_evol_+}) and (\ref%
{one_evol_-}), respectively. Hence the QRT holds in this case.

{\em {Case 2:}} $[A,L]=0$ or $[B,\tilde{L}(t)]=0$. In this case, using Eq.~(%
\ref{2time_evol_eq_f}), we obtain 
\begin{eqnarray}
d\langle \sigma _{-}(t_{1})\sigma _{z}(t_{2})\rangle /dt_{1} &=&-i\omega
_{A}\langle \sigma _{-}(t_{1})\sigma _{z}(t_{2})\rangle  \nonumber \\
&&-\left[ \Gamma _{1}(t_{1})+\Gamma _{2}^{\ast }(t_{1})\right] \langle
\sigma _{-}(t_{1})\sigma _{z}(t_{2})\rangle  \nonumber \\
&&-2\Gamma _{3}\left( t_{1},t_{2}\right) \left\langle \sigma
_{z}(t_{1})\sigma _{-}(t_{2})\right\rangle ,  \label{2time_evol_-z} \\
d\langle \sigma _{z}(t_{1})\sigma _{-}(t_{2})\rangle /dt_{1} &=&-\left[
\Gamma _{1}(t_{1})+\Gamma _{1}^{\ast }(t_{1})+\Gamma _{2}^{\ast
}(t_{1})+\Gamma _{2}(t_{1})\right]  \nonumber \\
&&\quad\quad \times \langle \sigma _{z}(t_{1})\sigma
_{-}(t_{2})\rangle  \nonumber \\
&&-\left[ \Gamma _{1}(t_{1})+\Gamma _{1}^{\ast }(t_{1})-\Gamma _{2}^{\ast
}(t_{1})-\Gamma _{2}(t_{1})\right] \nonumber \\
&&\quad\quad \times \langle \sigma
_{-}(t_{2})\rangle  \nonumber \\
&&-2\Gamma _{4}\left( t_{1},t_{2}\right) \left\langle \sigma
_{-}(t_{1})\sigma _{z}(t_{2})\right\rangle ,  \label{2time_evol_z-}
\end{eqnarray}%
where 
\begin{eqnarray}
\Gamma _{3}\left( t_{1},t_{2}\right) &=&\int_{0}^{t_{2}}d\tau \alpha
(t_{1}-\tau )e^{i\omega _{A}\left( t_{2}-\tau \right) },  \label{Gamma_3} \\
\Gamma _{4}\left( t_{1},t_{2}\right) &=&\int_{0}^{t_{2}}d\tau \beta
(t_{1}-\tau )e^{-i\omega _{A}\left( t_{2}-\tau \right) }.  \label{Gamma_4}
\end{eqnarray}%
When we obtain Eq.~(\ref{2time_evol_-z}) from Eq.~(\ref{2time_evol_eq_f}), the last term containing $\Gamma _{3}(t_{1},t_{2})$ in Eq.~(\ref{2time_evol_-z}) does not vanish since $[L^{\dagger },A]\neq 0$ and $[B,\tilde{L}(\tau -t_{2})]\neq 0$. Similarly, when we obtain Eq.~(\ref{2time_evol_z-}) from Eq.~(\ref{2time_evol_eq_f}), because $[L,A]\neq 0$ and 
$[B,\tilde{L}^{\dagger }(\tau -t_{2})]\neq 0$, the last term containing 
$\Gamma _{4}(t_{1},t_{2})$ in Eq.~(\ref{2time_evol_z-}) exists. Thus,
compared with single-time equations (\ref{one_evol_-}) and (\ref{one_evol_z}), 
the two-time equations (\ref{2time_evol_-z}) and (\ref{2time_evol_z-})
have the extra last terms containing $\Gamma _{3}(t_{1},t_{2})$ and 
$\Gamma _{4}(t_{1},t_{2})$, respectively. As a result, the QRT does not hold
in this case. It is also obvious from the sole appearance of the individual
coefficient of either $\Gamma _{3}(t_{1},t_{2})$ in 
Eq.~(\ref{2time_evol_-z}) or $\Gamma _{4}(t_{1},t_{2})$ 
in Eq.~(\ref{2time_evol_z-})
that the finite-temperature bath CF's $\alpha (t_{1}-\tau )$ and $\beta
(t_{1}-\tau )$ can not be combined into the single effective bath CF $\alpha
_{\mathrm{eff}}(t_{1}-\tau )=\alpha (t_{1}-\tau )+\beta (t_{1}-\tau )$ of
Eq.~(\ref{CFalpha_eff}) as in the Hermitian coupling operator case.

{\em {Case 3:}} $[A,L^{\dagger }]=0$ or $[B,\tilde{L}^{\dagger }(t)]=0$.
Using Eq.~(\ref{2time_evol_eq_f}), we obtain for this case 
\begin{eqnarray}
d\langle \sigma _{+}(t_{1})\sigma _{z}(t_{2})\rangle /dt_{1} &=&+i\omega
_{A}\langle \sigma _{+}(t_{1})\sigma _{z}(t_{2})\rangle  \nonumber \\
&&-\left[ \Gamma _{1}^{\ast }(t_{1})+\Gamma _{2}(t_{1})\right] \langle
\sigma _{+}(t_{1})\sigma _{z}(t_{2})\rangle  \nonumber \\
&&-2\Gamma _{4}\left( t_{1},t_{2}\right) \left\langle \sigma
_{z}(t_{1})\sigma _{+}(t_{2})\right\rangle ,  \label{2time_evol_+z} \\
d\langle \sigma _{z}(t_{1})\sigma _{+}(t_{2})\rangle /dt_{1} &=&-\left[
\Gamma _{1}(t_{1})+\Gamma _{1}^{\ast }(t_{1})+\Gamma _{2}^{\ast
}(t_{1})+\Gamma _{2}(t_{1})\right] \nonumber \\
&&\quad\quad \times \langle \sigma _{z}(t_{1})\sigma
_{+}(t_{2})\rangle  \nonumber \\
&&-\left[ \Gamma _{1}(t_{1})+\Gamma _{1}^{\ast }(t_{1})-\Gamma _{2}^{\ast
}(t_{1})-\Gamma _{2}(t_{1})\right] \nonumber \\
&&\quad\quad \times \langle \sigma
_{+}(t_{2})\rangle  \nonumber \\
&&-2\Gamma _{3}\left( t_{1},t_{2}\right) \left\langle \sigma
_{+}(t_{1})\sigma _{z}(t_{2})\right\rangle.  
\label{2time_evol_z+}
\end{eqnarray}
Similarly, compared with the single-time evolution equations 
(\ref{one_evol_+}) and (\ref{one_evol_z}), 
the two-time evolution equations (\ref{2time_evol_+z}) 
and (\ref{2time_evol_z+}) have the extra last terms
containing $\Gamma _{4}^{\ast }(t_{1})$ and $\Gamma _{3}(t_{1})$,
respectively. As a result, the QRT also does not hold for these two CF's.

{\em {Case 4:}} $[A,L]\neq 0$, $[A,L^{\dagger }]\neq 0$ and 
$[B,\tilde{L}(t)]\neq 0$ , $[B,\tilde{L}^{\dagger }(t)]\neq 0$. 
In this case, by using Eq.~(\ref{2time_evol_eq_f}), we obtain
 the following equations 
\begin{eqnarray}
d\langle \sigma _{z}(t_{1})\sigma _{z}(t_{2})\rangle /dt_{1} &=&-\left[
\Gamma _{1}(t_{1})+\Gamma _{1}^{\dagger }(t_{1})+\Gamma _{2}^{\dagger
}(t_{1})+\Gamma _{2}(t_{1})\right] \nonumber \\
&&\quad\quad \times \langle \sigma _{z}(t_{1})\sigma
_{z}(t_{2})\rangle   \nonumber \\
&&-\left[ \Gamma _{1}(t_{1})+\Gamma _{1}^{\dagger }(t_{1})-\Gamma
_{2}^{\dagger }(t_{1})-\Gamma _{2}(t_{1})\right] \nonumber \\
&&\quad\quad \times \langle \sigma
_{z}(t_{2})\rangle   \nonumber \\
&&+4\Gamma _{3}\left( t_{1},t_{2}\right) \left\langle \sigma
_{+}(t_{1})\sigma _{-}(t_{2})\right\rangle   \nonumber \\
&&+4\Gamma _{4}\left( t_{1},t_{2}\right) \left\langle \sigma
_{-}(t_{1})\sigma _{+}(t_{2})\right\rangle.  
\label{2time_evol_zz}
\end{eqnarray}
The evolution equation of the CF $\langle \sigma _{z}(t_{1})\sigma
_{z}(t_{2})\rangle $, Eq.~(\ref{2time_evol_zz}), is coupled with the evolution
equations of the CF's $\langle \sigma _{+}(t_{1})\sigma _{-}(t_{2})\rangle $
and $\langle \sigma _{-}(t_{1})\sigma _{+}(t_{2})\rangle $, which correspond
to the CF's in Case 2 and Case 3, respectively. Their evolution equations,
obtained from Eq.~(\ref{2time_evol_eq_f}), are 
\begin{eqnarray}
d\langle \sigma _{-}(t_{1})\sigma _{+}(t_{2})\rangle /dt_{1} &=&-i\omega
_{A}\langle \sigma _{-}(t_{1})\sigma _{+}(t_{2})\rangle   \nonumber \\
&&-\left[ \Gamma _{1}(t_{1})+\Gamma _{2}^{\ast }(t_{1})\right] \langle
\sigma _{-}(t_{1})\sigma _{+}(t_{2})\rangle   \nonumber \\
&&+\Gamma _{3}\left( t_{1},t_{2}\right) \left\langle \sigma
_{z}(t_{1})\sigma _{z}(t_{2})\right\rangle,   
\label{2time_evol_-+} \\
d\langle \sigma _{+}(t_{1})\sigma _{-}(t_{2})\rangle /dt_{1} &=&i\omega
_{A}\langle \sigma _{+}(t_{1})\sigma _{-}(t_{2})\rangle   \nonumber \\
&&-\left[ \Gamma _{1}^{\ast }(t_{1})+\Gamma _{2}(t_{1})\right] \langle
\sigma _{+}(t_{1})\sigma _{-}(t_{2})\rangle   \nonumber \\
&&+\Gamma _{4}\left( t_{1},t_{2}\right) \left\langle \sigma
_{z}(t_{1})\sigma _{z}(t_{2})\right\rangle.   
\label{2time_evol_+-}
\end{eqnarray}
From Eqs.~(\ref{2time_evol_zz}), (\ref{2time_evol_-+}) and (\ref%
{2time_evol_+-}), it is obvious that the QRT also does not hold for CF's $%
\langle \sigma _{z}(t_{1})\sigma _{z}(t_{2})\rangle $, $\langle \sigma
_{+}(t_{1})\sigma _{-}(t_{2})\rangle $ and $\langle \sigma _{-}(t_{1})\sigma
_{+}(t_{2})\rangle $.

\begin{figure}[tbp]
\includegraphics[width=0.95\linewidth]{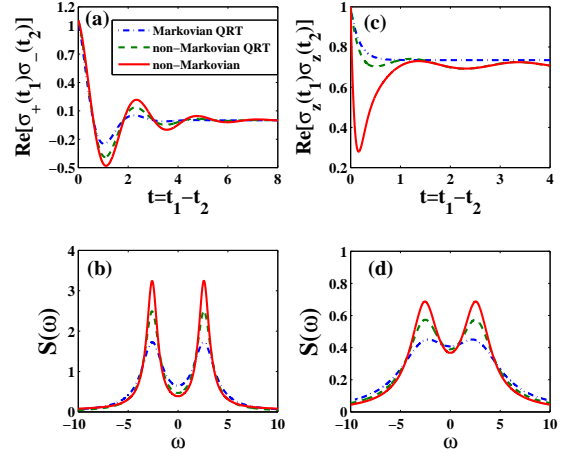}
\caption{(Color online) (a) Real part of the time evolution and (b) Fourier spectrum $S(\omega)$ of the system operator CF $\langle\sigma_+(t_1)\sigma_-(t_2)\rangle$, and (c) real part of 
$\langle \sigma _{z}(t_{1})\sigma _{z}(t_{2})\rangle $ for three different 
cases: Markovian using the QRT (blue dot-dashed line), non-Markovian
using the QRT (green dashed line) and non-Markovian (red solid line) using the
evolution equation (\ref{2time_evol_eq_f}). 
Other parameters used are $\omega_A=3$, $(k_BT/\hbar)=1$, $\Lambda=5$, $\gamma=0.1$, and $t_{2}=1$.
(d) Fourier spectrum $S(\omega)$ of $\langle\sigma_+(t_1)\sigma_-(t_2)\rangle$ for a different parameter of $\gamma =0.35$ and for an initial mixed system state.
The results of the non-Markovian QRT case and the non-Markovian
evolution case in (c) become indistinguishable when $t=t_1-t_2$ is larger
than 1.5.
}
\label{fig:CF}
\end{figure}

We may consider any spectral density for which the time-convolutionless
perturbation theory is still valid to characterize the environment, but for simplicity we consider a spectral density $J(\omega)=\sum_\lambda|g_\lambda|^2\delta(\omega-\omega_\lambda)=\gamma \hbar\omega (\omega/\Lambda)^{n-1}\exp(-\omega^2/\Lambda^2)$ with $n=1$ (Ohmic), where $\Lambda$ is the cut-off frequency and $\gamma$ is a dimensionless constant characterizing the interaction strength 
to the environment.      
Figure \ref{fig:CF} shows the real part of the time evolution 
of the system operator CF's $\langle\sigma_+(t_1)\sigma_-(t_2)\rangle$ and  
$\langle \sigma _{z}(t_{1})\sigma _{z}(t_{2})\rangle $, as well as 
the Fourier spectrum $S(\omega)$ of $\langle\sigma_+(t_1)\sigma_-(t_2)\rangle$.  
The CF's are obtained in three different cases: the first is in the
Markovian case [i.e, taking the reservoir CF's 
$\alpha(t_1-\tau)$ and 
$\beta(t_1-\tau)$ in Eq.~(\ref{2time_evol_eq_f}) to be $\delta$-correlated in time, or equivalently
taking the coefficients of $\Gamma_1$, $\Gamma_1^\dagger$, $\Gamma_2$,
and $\Gamma_2^\dagger$ to be time-independent and equal to their
Markovian long-time values and setting all $\Gamma_3(t_1,t_2)$ and
$\Gamma_4(t_1,t_2)$ to be zero in Eqs.~(\ref{2time_evol_zz}),
(\ref{2time_evol_-+}) and (\ref{2time_evol_+-})], the second is in the
non-Markovian case with a finite cut-off frequency but wrongly
directly using the QRT method [i.e., 
the last two terms of Eq.~(\ref{2time_evol_eq_f}) or equivalently the terms containing $\Gamma_3(t_1,t_2)$ or $\Gamma_4(t_1,t_2)$ in 
Eqs.~(\ref{2time_evol_zz}), (\ref{2time_evol_-+}) and (\ref{2time_evol_+-})
being all neglected],
and the third is in the non-Markovian case with a finite cut-off
frequency [i.e., using the evolution equation (\ref{2time_evol_eq_f})
or equivalently Eqs.~(\ref{2time_evol_zz}), (\ref{2time_evol_-+}) and
(\ref{2time_evol_+-}) derived
in this paper]. 
The initial environmental state is in the thermal state and the system state in Fig.~\ref{fig:CF}(a)--(c) is arbitrarily chosen to be $\vert \Psi\rangle =\left( \frac{\sqrt{3}}{2}\left\vert e\right\rangle +\frac{1}{2%
}\left\vert g\right\rangle \right)$. 
We can see that there are considerable differences between the results 
obtained in the three different cases in Fig.~\ref{fig:CF}(a) and (b), 
and more significant differences can be observed 
in Fig.~\ref{fig:CF}(c) and (d). 
The oscillations of the CF's are more pronounced in the non-Markovian cases. 
In Fig.~\ref{fig:CF}(b), the coherent peaks of the Fourier spectrum 
centered at $\omega=\pm\omega_A$
are higher and the widths are narrower in the non-Markovian cases.
The CF $\langle \sigma _{z}(t_{1})\sigma
_{z}(t_{2})\rangle $ of the non-Markovian evolution equation case (in
red solid line) in
Fig.~\ref{fig:CF}(c) 
differs more from the CF's of the other two cases (in green dashed
line and in blue dot-dashed line) in the short-time regime  
than the CF $\langle\sigma_+(t_1)\sigma_-(t_2)\rangle$ of the
non-Markovian evolution equation case (in red solid line) in
Fig.~\ref{fig:CF}(a) does.    
This is because as compared to the evolution equation of
$\langle\sigma_+(t_1)\sigma_-(t_2)\rangle$ of
Eq.~(\ref{2time_evol_+-}), the evolution equation of 
$\langle \sigma _{z}(t_{1})\sigma_{z}(t_{2})\rangle $ of
Eq.~(\ref{2time_evol_zz}) has, in addition to a term proportional to
$\Gamma_4(t_1,t_2)$,   an extra correction term 
proportional to $\Gamma_3(t_1,t_2)$ over its QRT counterparts.
It is also found that generally the results of the non-Markovian QRT
case (in green dashed lines) approach those
of the non-Markovian evolution equation case (in red solid lines)
more closely than the results of the Markovian QRT case (in blue
dot-dashed lines) do.
Similar behaviors are also observed when the temperature is increased
or when the cut-off frequency $\Lambda$ is
increased.
The Markovian case can be recovered from the non-Markovian ones in the
limit when the cut-off frequency $\Lambda\to\infty$, in which the
three results coincide. 
For a  
larger $\gamma$ and for an initial mixed system state with the values of the off-diagonal density matrix elements being a quarter of those of the pure state $\vert \Psi\rangle$, the peak heights of $S(\omega)$ 
are lower as shown in Fig.~\ref{fig:CF}(d).
Furthermore in Fig.~\ref{fig:CF}(d), 
the two coherent peaks are still clearly visible in the non-Markovian cases, 
while the two peaks is barely visible in the Markovian case.
  
For the present spin-boson model with the system coupling operator
$L\neq L^\dagger$, the self-Hamiltonian of the spin does not commute
with the system coupling operator, i.e, $[H_S,L]\neq 0$, and the
environment coupling operator also does not commute with the
self-Hamiltonian of the environment, i.e., $[H_R, a_\lambda]\neq 0$. 
Thus the exact non-Markovian finite-temperature two-time CF's of
the present spin-boson model are not directly available. 
But in Ref.~\cite{Goan10}, we evaluated the exact non-Markovian
finite-temperature two-time CF's of the system operators for an
exactly solvable pure-dephasing spin-boson model in two ways, one by
exact direct operator technique without any approximation and the
other by the derived evolution equation (\ref{2time_evol_eq_f})
valid to second order in the system-environment interaction Hamiltonian.
The perfect agreement of the results between the non-Markovian
evolution equation case and the exact operator evaluation case, and
the significant difference between the non-Markovian evolution
equation case and the case of wrongly applying non-Markovian QRT
\cite{Goan10}. 
demonstrate clearly the validity of the derived evolution equation
(\ref{2time_evol_eq_f}).
It is thus believed that in the weak system-environment coupling
limit, the finite-temperature CF's calculated
using our evolution equation that takes into account the nonlocal
environment memory term, Eq.~(\ref{non-Markovian_1st_order}), for the
present spin-boson model would
agree more closely with the exact 
results than those in the non-Markoian QRT and Markovian QRT
cases.

\section{Conclusion}\label{sec:conclusion}
In summary, we have derived evolution equations of the single-time and
two-time CF's of system operators, using a quantum master equation  
technique different from 
those presented in Refs.~\cite{Alonso05,Vega06,Alonso07}. 
This quantum master equation approach allows us to explicitly point
out an important nonlocal environment (bath) memory term that vanishes
in the Markovian case but makes the evolution equation deviate from
the QRT in general cases.   
The derived two-time 
equations are valid for thermal environments at any temperature with
Hermitian or non-Hermitian coupling operators and for any initially
factorized (separable) system-reservoir state (pure or mixed) as long as the  
assumption of Eq.~(\ref{traceless_1st_order}) and the approximation 
of the weak system-environment coupling 
that are used to derive the equations apply. 
In contrast to the evolution equations presented in
Refs.~\cite{Alonso05,Vega06,Alonso07,Goan10} that are applicable for
bosonic environments, Eq.~(\ref{2time_evol_eq}) derived in this paper
can be used to calculate the two-time CF's for a wide range of 
system-environment models with
bosonic and/or fermionic environments.
We have also given conditions on which the QRT holds in the weak
system-environment coupling case and have applied the derived equations to
a problem of a two-level system (atom) coupled to a
finite-temperature 
thermal bosonic environment (electromagnetic fields),
in which the system coupling operator is not Hermitian, $L\neq
L^\dagger$, and the evolution equations derived in
Refs.~\cite{Alonso05,Vega06,Alonso07} are not applicable.  
It is easy to calculate the two-time CF's using the
derived evolution equations. Other non-Markovian open quantum system
models that are not exactly solvable can be proceeded in a similar way
to obtain the time evolutions of their two-time system operator CF's
valid to second order in the system-environment interaction
Hamiltonian. This illustrates the practical usage of the evolution
equations. 
Therefore, the derived evolution equations that
generalize the QRT to the non-Markovian cases will
have broad applications in
many different branches of physics, such as quantum optics,
statistical physics, chemical physics, quantum
transport in nanostructure devices and so forth when the properties
related to the two-time CF's are of interests. 

\begin{acknowledgments}
We would like to acknowledge support from the National Science
Council, Taiwan, under Grant No. 97-2112-M-002-012-MY3, 
support from the Frontier and Innovative Research Program 
of the National Taiwan University under Grants No. 99R80869 and 
No. 99R80871, and support from the focus group program of the National Center for Theoretical Sciences, 
Taiwan. H.S.G. is grateful to the National Center for High-performance Computing, Taiwan, 
for computer time and facilities.
\end{acknowledgments}


\begin{thebibliography}{99}
\bibitem{direct_product} In Ref.~\cite{Goan10}, 
the subscript of ${S\oplus R}$ attached to the
trace symbol of ${\rm Tr}$ was used to represent a trace over the
degrees of freedom of the total composite system. Here, a more
commonly used notation of ${S\otimes R}$ is adopted for a trace over
the direct-tensor-product Hilbert space of the system and the reservoir.
\end{thebibliography}
\end{document}